\documentclass[aps,prd,preprint]{revtex4-1}

\usepackage{graphicx}  % needed for figures
\graphicspath{ {images/} } %the images are kept in a folder named images under the current directory.
\usepackage{latexsym}
\usepackage{slashed}
\usepackage{dcolumn}   % needed for some tables
\usepackage{bm}        % for math
\usepackage{amsmath}
\usepackage{amssymb}   % for math
\usepackage{longtable}
\usepackage{float}
\newcommand{\be}{\begin{equation}}
\newcommand{\ee}{\end{equation}}
\newcommand{\bea}{\begin{eqnarray}}
\newcommand{\eea}{\end{eqnarray}}
\newcommand{\bma}{\begin{matrix}}
\newcommand{\ema}{\end{matrix}}
\newcommand{\nn}{\nonumber}
\newcommand{\bml}{\begin{mathletters}}
\newcommand{\eml}{\end{mathletters}}
\newcommand{\bes}{\begin{subequations}}
\newcommand{\ees}{\end{subequations}}

\newcommand{\bi}{\begin{itemize}}
\newcommand{\ei}{\end{itemize}}
\newcommand{\gev}{~{\rm GeV}}

\begin{document}
\title{On neutrino and charged lepton masses and mixings: A view from the electroweak-scale right-handed neutrino model }
\author{P. Q. Hung}
\email{pqh@virginia.edu}
\author{Trinh Le}
\email{ttl9ve@virginia.edu}
\affiliation{Department of Physics, University of Virginia,
Charlottesville, VA 22904-4714, USA}

\date{\today}

\begin{abstract}
We present a model of neutrino masses within the framework of the EW-$\nu_R$ model in which the experimentally desired form of the PMNS matrix is obtained by applying an $A_4$ symmetry to the \emph{Higgs singlet sector} responsible for the neutrino Dirac mass matrix. This mechanism naturally avoids potential conflict with the LHC data which severely constrains the Higgs sector, in particular the Higgs doublets. Moreover, by making a simple $ans\ddot{a}tz$ we extract $\mathcal{M}_l {\mathcal{M}_l}^\dagger$ for the charged lepton sector. A similar $ans\ddot{a}tz$ is proposed for the quark sector. The sources of masses for the neutrinos are entirely different from those for the charged leptons and for the quarks and this might explain why $U_{PMNS}$ is {\em very different} from $V_{CKM}$.
\end{abstract}

\pacs{}\maketitle

\section{Introduction}
The discovery and subsequent analyses of neutrino oscillation phenomena have revealed a trove of valuable information concerning the mixing matrix $U_{PMNS}$ and the mass difference squared in the neutrino sector. It is worth stressing that the mere presence of neutrino masses as implied by the oscillation data \cite{Nakamura:2010zzi} provides the first evidence of physics beyond the Standard Model (BSM). What might be the origin of neutrino masses? Why are they so tiny ($m_{\nu} < O(eV)$) as compared with even that of the lightest of elementary particles: the electron?  Why is the leptonic mixing matrix $U_{PMNS}$ so different from $V_{CKM} $ of the quark sector? Is there any chance that some of the physics that are responsible for the tininess of the neutrino masses as well as their mixings could somehow be experimentally accessible at the Large Hadron Collider (LHC) in the near future or even at the International Linear Collider at a not-too-distant future?

The vast difference between neutrino masses and those of other elementary particles is a big mystery. There could be several ways in which neutrinos can obtain masses, all of which go beyond the Standard Model. The most obvious one is to add right-handed neutrinos which are singlets of the SM and couple them through Yukawa interactions with the left-handed lepton doublets and the SM Higgs doublet. In this simplest Dirac mass scenario, the Yukawa coupling would have to be {\em unnaturally} small i.e. $g_{\nu} \lesssim O(10^{-11})$ in order to accommodate $m_{\nu} < O(eV)$. A more elegant scenario is the quintessential see-saw mechanism \cite{seesaw} where the right-handed neutrinos acquire a Majorana mass term $M_R \nu_{R}^T \sigma_2 \nu_R$ in addition to a Dirac mass term $m_D \nu_{L}^\dagger \nu_{R} + H.c.$. The diagonalization of the mass matrix yields two eigenvalues whose magnitudes are approximately $m_{D}^2/M_R$ and $M_R$ for $M_R \gg m_D$. In a generic see-saw scenario, right-handed neutrinos are SM singlets (e.g. in an SO(10) scenario) i.e. they are {\em sterile}, and typically $m_D \propto O(\Lambda_{EW})$ and $M_R \propto O(\Lambda_{GUT})$ or $O(M_{W_R})$. Although this generic scenario can elegantly "explain" the smallness of neutrino masses,  it goes without saying that the prospect of directly testing the seesaw mechanism by searching for right-handed neutrinos is very remote, both from an energetic point of view and from a production point of view. (Although it is very popular, leptogenesis by itself is not such a direct test.)

One of us (PQH) has proposed a model \cite{pqnur} of electroweak-scale right-handed neutrinos in which (as we will briefly review below) $\nu_R$'s belong to $SU(2)$ doublets along with mirror charged leptons. This has two distinct advantages: 1) $\nu_R$'s are {\em non-sterile} and couple to the Z and W bosons; 2) Since $\nu_R$'s  are members of doublets, a Majorana mass term necessarily comes from the vacuum expectation value (VEV) of a triplet Higgs field which spontaneously breaks $SU(2) \times U(1)_Y$ (in addition to the Higgs doublet) and, as a result, $M_R \propto O(\Lambda_{EW})$. In this scenario, the EW $\nu_R$ model, right-handed neutrinos can be produced and searched for at the LHC or at the proposed ILC. 

The EW $\nu_R$ model  \cite{pqnur} keeps the same gauge group as the SM i.e. $SU(3)_C \times SU(2) \times U(1)_Y$ but increases its fermion as well as its Higgs content, all of which are listed below. In a nutshell, for every left-handed fermion doublet there is a right-handed mirror fermion doublet, and for every right-handed fermion $SU(2)$ singlet, there is a left-handed mirror fermion singlet. On the scalar sector side, the minimal EW $\nu_R$ model contains one Higgs doublet, one complex triplet, one real triplet and one Higgs singlet. The role of these Higgs fields in generating fermion masses will be discussed below. In particular, we will discuss the importance of the Higgs singlet on the issue of neutrino masses.

Before moving on to the main discussion of this paper, namely $U_{PMNS}$ and its implications for fermion masses, a few remarks concerning the viability of the model concerning the electroweak precision test and in light of the 125-GeV SM-like Higgs boson are in order.
It turns out that the EW $\nu_R$ model  passes the electroweak precision data test very well as shown in \cite{hung2} due to the presence of the Higgs triplets. The positive contributions to the S-parameter from the mirror fermions get partially cancelled by those coming from the Higgs triplets. As for the 125-GeV SM-like Higgs boson, the minimal EW $\nu_R$ model contains a CP-odd state that could, in principle, be that 125-GeV candidate. However, a likelihood analysis \cite{cms} indicated that the 125-GeV SM-like Higgs boson is more likely to be a CP-even state. A minimal extension of the EW $\nu_R$ model to include a second Higgs doublet has revealed an interesting aspect of the nature of the 125-GeV object which is extensively discussed in \cite{hung3}.

The plan of the paper will be as follows. First we give a brief summary of the EW $\nu_R$ model, in particular its Yukawa sector. Next, we present a summary of the results obtained in \cite{hung2} and \cite{hung3}. We then discuss the motivation for using the non-Abelian discrete symmetry $A_4$ \cite{Ishimori:2010au} to describe the Dirac part of the neutrino mass matrix which, in the EW $\nu_R$ model, is generated by the Higgs singlets. In this paper, we increase the number of Higgs singlets from {\em one} (the number in the original model) to {\em four} without any consequence as far as the 125-GeV SM-like Higgs boson is concerned. The number of non-singlet Higgs fields is kept unchanged in view of the tight constraints coming  from the properties of the 125-GeV object as discussed in \cite{hung3}. 

\section{A brief review of the EW $\nu_R$ model \cite{pqnur}} 
%\subsection{The EW$\nu_R$ Model \cite{pqnur}}

\subsection{{\bf Gauge group}}

The gauge group of the EW $\nu_R$ model stays the same as that of the SM, namely:

\be
SU(3)_C \times SU(2) \times U(1)_Y \,. \nonumber
\ee

\subsection{{\bf Particle Content}}
{\bf Leptons and quarks} (generic notations):

\begin{itemize}
\item Doublets
	\begin{itemize}
	 \item SM: $l_L = \left(
	  \begin{array}{c}
	   \nu_L \\
	   e_L \\
	  \end{array}
	 \right); \
	 	q_L = \left(
	  	 \begin{array}{c}
	   	  u_L \\
	     	  d_L \\
	  	\end{array}
	 	\right)$
	 \item Mirror: $l_R^M = \left(
	  \begin{array}{c}
	   \nu_R^M \\
	   e_R^M \\
	  \end{array}
	 \right); \
	 	q_R^M = \left(
	  	 \begin{array}{c}
	   	  u_R^M \\
	     	  d_R^M \\
	  	\end{array}
	 	\right)$
	\end{itemize}
\item Singlets
	\begin{itemize}
		 \item SM: $e_R; \ u_R, \ d_R$
 		\item Mirror: $e_L^M; \ u_L^M, \ d_L^M$
	\end{itemize}
\end{itemize}

{\bf Higgs fields}:
	\begin{itemize}
	 \item A singlet scalar Higgs $\phi_S$ with $\langle \phi_S \rangle = v_S$.
	 \item Doublet Higgses:
	 
	  $\Phi_2=\left(
	 			\begin{array}{c}
				\phi_{2}^+ \\
				\phi_{2} ^0 \\
				\end{array}
				\right)$ 		
with $\langle \phi_{2}^0 \rangle = v_2/\sqrt{2}$. 

In the original version \cite{pqnur}, this Higgs doublet couples to both SM and mirror fermions. An extended version was proposed \cite{hung2} in order to accommodate the 125-GeV SM-like scalar and, in this version, $\Phi_2$ only couples to SM fermions while another doublet $\Phi_{2M}$ whose VEV is $\langle \phi_{2M}^0 \rangle = v_{2M}/\sqrt{2}$ couples only to mirror fermions. Since this paper focuses only on SM fermions, we will concentrate only on $\Phi_2$. 

	 \item Higgs triplets
	 	\begin{itemize}
			\item $\widetilde{\chi} \ (Y/2 = 1)  = \frac{1}{\sqrt{2}} \ \vec{\tau} . \vec{\chi} = 
	  \left(
	  \begin{array}{cc}
	    \frac{1}{\sqrt{2}} \chi^+ & \chi^{++} \\
	    \chi^0 & - \frac{1}{\sqrt{2}} \chi^+\\
	   \end{array}
		  \right)$ with $\langle \chi^0 \rangle = v_M$.
			\item $\xi \ (Y/2 = 0)$ in order to restore Custodial Symmetry with $\langle \xi^0 \rangle = v_M$.
			
	\item VEVs:
	
	$v_{2}^2 +v_{2M}^2 + 8\,v_{M}^2= v^2 \approx (246 \gev)^2$  		
		\end{itemize}		  	 
	\end{itemize}

\subsection{{\bf Dirac and Majorana Neutrino Masses}}
For simplicity, from hereon, we will write $\nu_R^M$ simply as $\nu_R$.
\begin{itemize}
\item Dirac Neutrino Mass\\
The singlet scalar field $\phi_S$ couples to fermion bilinear
\bea
\label{dirac}
L_S &=& g_{Sl} \,\bar{l}_L \ \phi_S \ l_R^M + h.c.\\  \nonumber
         &= &g_{Sl} (\bar{\nu}_L \ \nu_R \ + \bar{e}_L \ e_R^M) \ \phi_S + h.c. \,.
\eea
From (\ref{dirac}),  we get the Dirac neutrino masses $m_\nu^D = g_{Sl} \ v_S $.\\
\item Majorana Neutrino Mass
\bea
\label{majorana}
L_M &= &g_M \, l^{M,T}_R \ \sigma_2 \ \tau_2 \ \tilde{\chi} \ l^M_R \\ \nonumber
&= &g_M \ \nu_R^T \ \sigma_2 \ \nu_R \ \chi^0 - \dfrac{1}{\sqrt{2}} \ \nu_R^T \ \sigma_2 \ e_R^M \ \chi^+ \\ \nonumber
&&- \dfrac{1}{\sqrt{2}} \ e_R^{M,T} \ \sigma_2 \ \nu_R \ \chi^+ + e_R^{M,T} \ \sigma_2 \ e_R^M \ \chi^{++} \,.
\eea
From (\ref{majorana}), we obtain the Majorana mass $ M_R = g_M v_M $.
\end{itemize}

\section{Review of results of the EW $\nu_R$ model as discussed in \cite{hung2} and \cite{hung3}}

In this review section, we will discuss two sets of results for the EW $\nu_R$ model obtained in \cite{hung2} (the electroweak precision constraints) and \cite{hung3} (constraints from the 125-GeV SM-like scalar). 

\subsection{Electroweak precision constraints on the EW $\nu_R$ model \cite{hung2}}

The presence of mirror quark and lepton $SU(2)$-doublets can, by themselves, seriously affect the constraints coming from electroweak precision data. As noticed in \cite{pqnur}, the positive contribution to the S-parameter coming from the extra right-handed mirror quark and lepton doublets could be partially cancelled by the negative contribution coming from the triplet Higgs fields. Ref.~\cite{hung2} has carried out a detailed analysis of the electroweak precision parameters S and T and found that there is a large parameter space in the model which satisfies the present constraints and that there is {\em no fine tuning} due to the large size of the allowed parameter space. It is beyond the scope of the paper to show more details here but a representative plot would be helpful. Fig. 1 shows the contribution of the scalar sector versus that of the mirror fermions to the S-parameter within 1$\sigma$ and 2$\sigma$.
\begin{figure}[H]
\centering
    \includegraphics[scale=0.35]{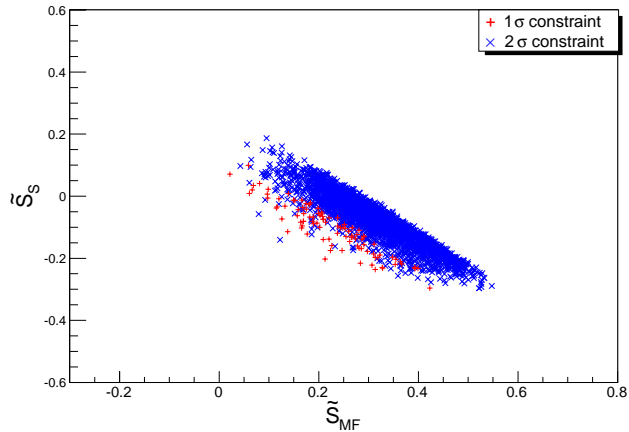} 
 \caption{{\small Constrained $\tilde{S}_S$ versus $\tilde{S}_{MF}$}}
\label{SsvsSmf}
\end{figure}
In the above plot, \cite{hung2} took for illustrative purpose 3500 data points that fall inside the 2$\sigma$ region with about 100 points falling inside the 1$\sigma$ region. More details can be found in \cite{hung2}.

\subsection{Review of the scalar sector of the EW $\nu_R$ model in light of the discovery of the 125-GeV SM-like scalar \cite{hung3}}

In light of the discovery of the 125-GeV SM-like scalar, it is imperative that any model beyond the SM (BSM) shows a scalar spectrum that contains at least one Higgs field with the desired properties as required by experiment. The present data from CMS and ATLAS only show signal strengths that are compatible with the SM Higgs boson. The definition of a signal strength $\mu$ is as follows 
\be
\sigma(H \text{-decay}) = \sigma(H \text{-production}) \times BR(H \text{-decay})\,,
\ee
and
\be\label{eq:mudef}
	\mu(H \text{-decay}) = \frac{\sigma(H \text{-decay})}{\sigma_{SM}(H \text{-decay})}\,.
\ee

To really distinguish the SM Higgs field from its impostor, it is necessary to measure the partial decay widths and the various branching ratios. In the present absence of such quantities, the best one can do is to present cases which are consistent with the experimental signal strengths. This is what was carried out in \cite{hung3}. 

The minimization of the potential containing the scalars shown above breaks its global symmetry $SU(2)_L \times SU(20_R$ down to a custodial symmetry $SU(2)_D$ which guarantees at tree level $\rho = M_{W}^2/M_{Z}^2 \cos^2 \theta_W=1$ \cite{hung3}. The physical scalars can be grouped, based on their transformation properties under $SU(2)_D$ as follows:
	\begin{eqnarray}
		\text{five-plet (quintet)} &\rightarrow& H_5^{\pm\pm},\; H_5^\pm,\; H_5^0;\nonumber\\[0.5em]
		\text{triplet} &\rightarrow& H_{3}^\pm,\; H_{3}^0;\nonumber\\[0.5em]
		\text{triplet} &\rightarrow& H_{3M}^\pm,\; H_{3M}^0;\nonumber\\[0.5em]
		\text{three singlets} &\rightarrow& H_1^0,\; H_{1M}^0,\; H_1^{0\prime}\,,
	\end{eqnarray}
  The three custodial singlets are the CP-even states, one combination of which can be the 125-GeV scalar. In terms of the original fields, one has $H_1^0 = \phi_{2}^{0r}$,  $H_{1M}^0 = \phi_{2M}^{0r}$, and $H_1^{0\prime} = \frac{1}{\sqrt{3}} \Big(\sqrt{2}\chi^{0r}+ \xi^0\Big)$. These states mix through a mass matrix obtained from the potential and the mass eigenstates are denoted by $\widetilde{H}$, $\widetilde{H}^\prime$, and $\widetilde{H}^{\prime\prime}$, with the convention that the lightest of the three is denoted by $\widetilde{H}$, the next heavier one by $\widetilde{H}^\prime$ and the heaviest state by $\widetilde{H}^{\prime\prime}$. 
  
  To compute the signal strengths $\mu$, Ref.~\cite{hung3} considers $\widetilde{H} \rightarrow ZZ,~W^+W^-,~\gamma\gamma,~b\bar{b},~\tau\bar{\tau}$. In addition, the cross section of $g g \rightarrow \widetilde{H}$ related to $\widetilde{H} \rightarrow g g$ was also calculated. A scan over the parameter space of the model yielded {\em two interesting scenarios} for the 125-GeV scalar: 1) {\em Dr Jekyll}'s scenario in which $\widetilde{H} \sim H_1^0$ meaning that the SM-like component $H_1^0 = \phi_{2}^{0r}$ is {\em dominant}; 2) 
{\em Mr Hyde}'s scenario in which $\widetilde{H} \sim H_1^{0\prime}$ meaning that the SM-like component $H_1^0 = \phi_{2}^{0r}$ is {\em subdominant}. Both scenarios give signal strengths compatible with experimental data as shown below in Fig.~(2).
\begin{figure}
\label{signal2}
	\centering
	\includegraphics[width=0.5\textwidth]{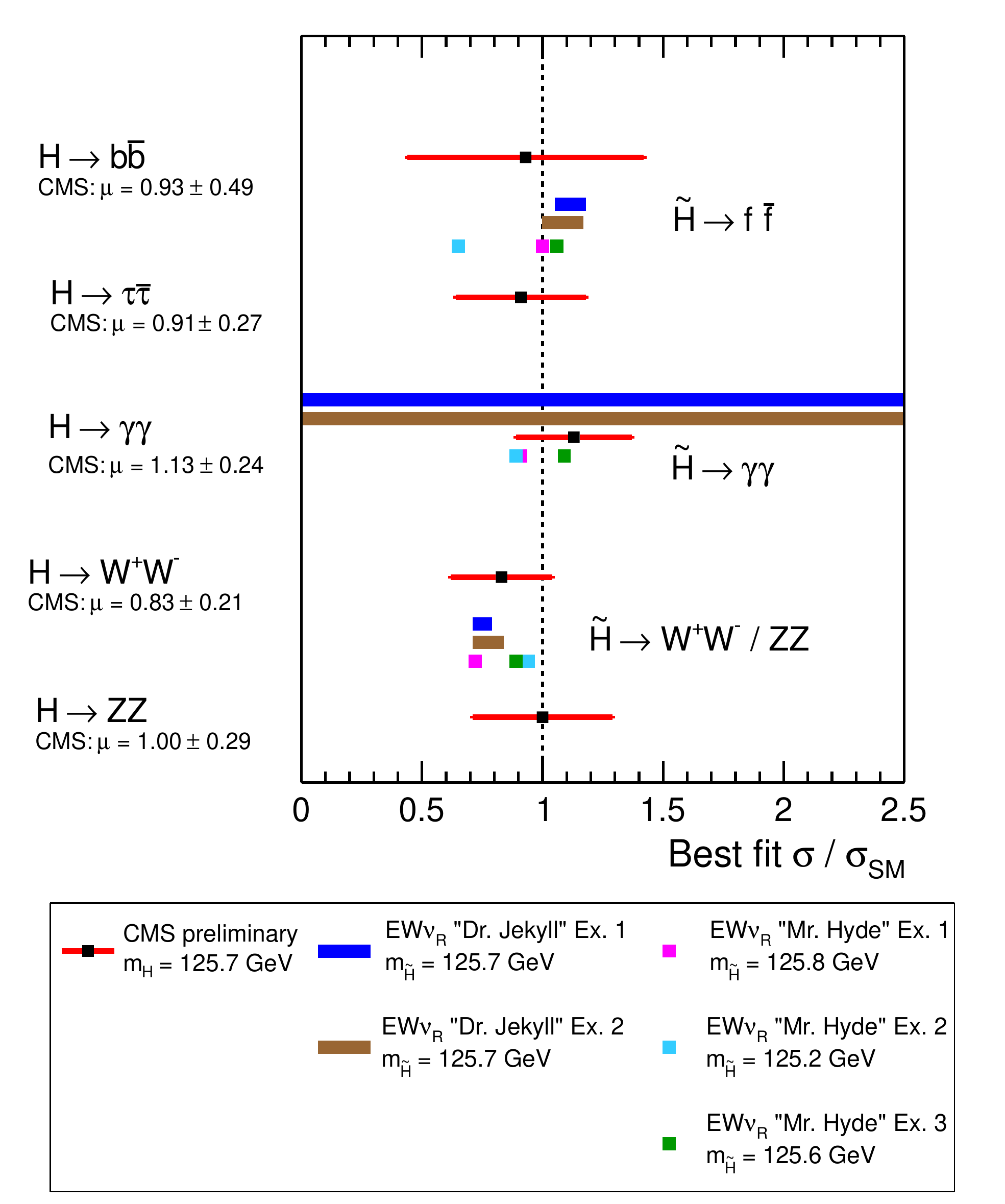}
	\caption{Figure shows the predictions of $\mu(\widetilde{H} \rightarrow ~b\bar{b}, ~\tau\bar{\tau}, ~\gamma\gamma, ~W^+W^-, ~ZZ)$ in the EW $\nu_R$ model for examples 1 and 2 in {\em Dr.~Jekyll} and example 1, 2 and 3 in {\em Mr.~Hyde} scenarios as discussed in \cite{hung3}, in comparison with corresponding best fit values by CMS \cite{h_ww_122013, h_zz_4l_122013, h_bb_102013, h_tautau_012014}.}
\end{figure}

As we can see from Fig.~(2), both SM-like scenario ({\em Dr Jekyll}) and the {\em more interesting scenario} which is very unlike the SM ({\em Mr Hyde}) agree with experiment. As stressed in \cite{hung3}, present data cannot tell whether or not the 125-GeV scalar is truly SM-like or even if it has a dominant SM-like component. It has also been stressed in \cite{hung3} that it is essential to measure the partial decay widths of the 125-GeV scalar to truly reveal its nature. Last but not least, in both scenarios, $H_{1M}^0 = \phi_{2M}^{0r}$ is subdominant but is essential to obtain the agreement with the data as shown in \cite{hung3}.

As discussed in detail in \cite{hung3} , for proper vacuum alignment, the potential contains a term proportional to $\lambda_5$ (Eq.~(32) of \cite{hung3}) and it is this term that prevents the appearance of Nambu-Goldstone (NG) bosons in the model. The would-be NG bosons acquire a mass proportional to $\lambda_5$ .

An analysis of CP-odd scalar states $H_{3}^0, H_{3M}^0 $ and the heavy CP-even states $\widetilde{H}^\prime$, and $\widetilde{H}^{\prime\prime}$ was presented in \cite{hung3}. The phenomenology of charged scalars including the doubly-charged ones was also discussed in \cite{pqaranda}.

The phenomenology of mirror quarks and leptons was briefly discussed in \cite{pqnur} and a detailed analysis of mirror quarks will be presented in \cite{oklahoma}. It suffices to mention here that mirror fermions decay into SM fermions through the process $q^M\rightarrow q\phi_S$, $l^M\rightarrow l\phi_S$ with $\phi_S$ "appearing" as missing energy in the detector. Furthermore, the decay of mirror fermions into SM ones can happen outside the beam pipe and inside the silicon vertex detector. Searches for non-SM fermions do not apply in this case. It is beyond the scope of the paper to discuss these details here.

This concludes the brief summary of the EW $\nu_R$ model \cite{pqnur}. The original minimal model contains just one singlet Higgs field $\phi_S$. As we shall see below, the $A_4$ symmetry that we will be using will necessitate an extension to {\em four} Higgs singlet fields with no phenomenological constraints at the present time. We will come back to this point below.

%-----------------------------------------------------------------------------
\section{Model of Neutrino Masses in the EW-$\nu_R$ model}

It is a big puzzle why the quark mixing matrix, the so-called CKM matrix \cite{Zhao:2012sqa}

\be
\lvert V_{CKM} \rvert =
\left(
  \begin{array}{ccc}
  0.9743 \pm 0.0002 & 0.2255 \pm 0.0024 & (5.10 \pm 0.47) \times 10^{-3} \\
  0.230 \pm 0.011 & 1.006 \pm 0.023 & (40.9 \pm 1.1) \times 10^{-3} \\
  (8.4 \pm 0.6) \times 10^{-3} & (42.9 \pm 2.6) \times 10^{-3} & 0.89 \pm 0.07 \\
  \end{array}
\right)
\ee
(which is not too different from the unit matrix)  differs so much from the leptonic one, the so-called Pontecorvo-Maki-Nakagawa-Sakata (PMNS) matrix \cite{Zhao:2012sqa}

\be
\label{exppmns}
\lvert U_{PMNS} \rvert = 
\left(
  \begin{array}{ccc}
  0.779...0.848 & 0.510...0.604 & 0.122...0.190 \\
  0.183...0.568 & 0.385...0.728 & 0.613...0.794 \\
  0.200...0.576 & 0.408...0.742 & 0.589...0.775 \\
  \end{array}
\right) \,.
\ee

Although the precise mass mechanism is far from being understood, it is not too unreasonable to speculate that the aforementioned big difference arises from the way neutrinos obtain masses as compared with the way charged fermions obtain theirs.
In this section, we will present a model in which it is the Dirac mass matrix of the neutrinos that is obtained by incorporating an $A_4$ symmetry into the model. In a nutshell, the $A_4$ group is a non-Abelian discrete group \cite{Ishimori:2010au} with four irreducible representations: $\underline{1}$, $\underline{1}^\prime$, $\underline{1}^{\prime \prime}$, $\underline{3}$, and with the product rule given by (using the notation of \cite{ma}):
\bea
\label{product}
\underline{3} \times \underline{3}& =& \underline{1}(11+22+33) + \underline{1}^\prime (11+\omega^2 22 + \omega 33) + \underline{1}^{\prime \prime}(11+\omega 22 + \omega^2 33)  \\ \nonumber 
&& + \underline{3} (23,31,12) + \underline{3} (32,13,21) \,.
\eea
Here 
\be
\label{omega}
\omega = e^{i2\pi/3}
\ee

Which particles are assigned to which representations of $A_4$ is a question which depends entirely on the model one is dealing with.  Below we briefly summarize two popular $A_4$ models in order to show the contrasts with ours.

$A_4$ has widely been used to produce the tribimaximal form of the PMNS matrix \cite{ma}. In a nutshell, as summarized nicely in \cite{ma}, the  $A_4$ symmetry is usually applied to the charged lepton mass matrix with the result being that the unitary matrix which diagonalizes the charged lepton mass matrix takes on the form of the PMNS that was first proposed by Cabibbo and Wolfenstein \cite{CW}, namely
\be
\label{CW}
U_{CW} = \frac{1}{\sqrt{3}}
\left(
  \begin{array}{cccc}
  1 & 1 & 1 \\
  1 & \omega & \omega^2 \\
  1 & \omega^2 & \omega \\
  \end{array}
\right)
\ee
In popular versions of $A_4$-inspired models of neutrino mass and mixing, $U_{CW}$ is identified with the unitary matrix $U_{lL}$ which diagonalizes the {\em charged lepton mass matrix}. 

In one version \cite{ma}, the left-handed lepton doublets are assigned to $\underline{3}$ while the right-handed $SU(2)$-singlet charged leptons are assigned to $\underline{1}, \underline{1}^\prime,  \underline{1}^{\prime \prime}$. There are three Higgs doublets belonging to $\underline{3}$. In another version, there are 4 Higgs doublets and the right-handed charged leptons belong to a $\underline{3}$. As described in \cite{ma}, the main feature of these models is the diagonalization of the {\em charged lepton} mass matrix by $U_{CW}$ in Eq.~(\ref{CW}) and the presence of three or four Higgs doublets. We shall make some remarks concerning the implication of the 125-GeV SM-like boson on models with extended Higgs sectors.

In  \cite{BMV}, a supersymmetric model was written which now includes three families of $SU(2)$-singlet vector-like heavy quarks and leptons, two Higgs doublets transforming as $\underline{1}$ and three Higgs singlets that transform as $\underline{3}$ of $A_4$. Here the SM right-handed fermions transform like $\underline{1}, \underline{1}^\prime,  \underline{1}^{\prime \prime}$. Just as with the models mentioned above \cite{ma}, the construction is such that $U_{CW}$ in Eq.~(\ref{CW}) is the matrix which diagonalizes the charged lepton mass matrix.

The two examples discussed above are two of several scenarios making use of the $A_4$ symmetry. It is beyond the scope of this paper to compare our approach with all others that are present in the literature. The main point we would like to stress here is that the most popular scenario is one in which the $A_4$ symmetry is used to generate $U_{CW}$ in Eq.~(\ref{CW}) for the {\em charged lepton sector}. Also, right-handed neutrinos in most generic models are SM-singlets and their Majorana masses are expected to be much larger than the EW scale.

Before discussing our approach based on $A_4$, we would like to point out the main differences with the aforementioned scenarios: 1) The conjugate of the matrix as shown in Eq.~(\ref{CW}) is the one that diagonalizes the {\em neutrino Dirac mass matrix}; 2) Right-handed neutrinos belong to $SU(2)$ doublets along with mirror charged leptons as espoused in \cite{pqnur} and are therefore non-sterile. Their Majorana masses are proportional to the EW symmetry breaking scale. 
%As the most mysterious particles in the world neutrinos have been studied for years by many physicists. The fact that neutrinos have masses motivates particle theorists to "create" an adequate model which leads them to a clear understanding of neutrino masses. The EW-$\nu_R$ model combing with the $A_4$ symmetry is a new approach to our big goal because it introduces three new scalar Higgs singlets $\phi_{1S}$, $\phi_{2S}$, $\phi_{3S}$ without conflicting with the LHC data which severely constrained the Higgs doublet sector. Moreover, the reason which makes the $A_4$ symmetry become a good candidate for this task is its simplicity and sufficiency. \\

Let us first start out with assignments of the EW $\nu_R$ model's content under $A_4$ %and L (lepton number).
%\vspace{1ex}
\begin{table}[!htb]
\caption{\label{assignment} $A_4$ assignments for leptons and Higgs fields}
\begin{center}
\begin{tabular}{ |m{1.2cm}||m{1.5cm}|m{1.5cm}|m{1cm}|m{1cm}|m{1cm}|m{1cm}|m{1cm}|m{1cm}|} 
%\begin{tabular}{| c || c | c | c | c | c | c | c |}
 \hline
 Field & $\mathnormal{(\nu,l)_L}$ & $\mathnormal{(\nu, l^M)_R}$ & $\mathnormal{e_R}$ & $\mathnormal{e_L^M}$ & $\mathnormal{\phi_{0S}}$ & $\mathnormal{\tilde{\phi}_{S}}$ & $\mathnormal{\Phi_2}$ \\ [0.5ex]
 \hline
 $A_4$ & $\underline{3}$ & \underline{3} & \underline{3} & \underline{3} & \underline{1} & \underline{3} & \underline{1}\\
  \hline
\end{tabular}
\end{center}
\end{table}

Notice that had the singlet Higgs fields belonged to $\underline{1}$, $\underline{1}^{'}$, and $\underline{1}^{''}$ only, the neutrino Dirac mass matrix would be diagonal which is not a desired scenario.
%==========
\subsection{Neutrino Dirac mass matrix}
As shown in \cite{pqnur}, the neutrino Dirac mass in the EW $\nu_R$ model comes from the generic Yukawa term $g_{Sl}  \, \bar{l}_{L}\, \phi_S \, l^{M}_{R} + H.c. $ (\ref{dirac}). With the  $A_4$ assignments shown in Table~(\ref{assignment}), we can write the following Yukawa interactions
\be
\label{yukawa}
L_S = \bar{l}_{L}\, (g_{0S} \phi_{0S} + g_{1S} \tilde{\phi}_S +  g_{2S} \tilde{\phi}_S )\, l^{M}_{R} + H.c. \,,
\ee
where $g_{1S}$ and $g_{2S}$ reflect the two different ways that $\tilde{\phi}_{S}$ couples to the product of $\bar{l}_{L}$ and $l^{M}_{R}$ as shown in Eq.~(\ref{product}).
We obtain the following neutrino Dirac mass matrix:
\be
\label{mnu}
M_\nu^D = 
\left(
  \begin{array}{cccc}
    g_{0S}v_0 & g_{1S}v_3 & g_{2S}v_2 \\
    g_{2S}v_3 & g_{0S}v_0 & g_{1S}v_1 \\
    g_{1S}v_2 & g_{2S}v_1 & g_{0S}v_0 \\
  \end{array}
\right) \, ,
\ee
where $v_0 =\langle \phi_{0S} \rangle$ and $v_i =\langle \phi_{iS} \rangle$ with $\imath=1,2,3$. Notice that this form of $M_\nu^D$ is the same as the one used by \cite{ma} for the {\em charged lepton mass matrix}. 

When $v_1=v_2=v_3=v$, $M_\nu^D$ can be diagonalized as follows (using $1+ \omega + \omega^2=0$ and $\omega^2=\omega^{\ast}$)
\be
\label{diagonal}
U_{\nu}^{\dagger} M_\nu^D U_{\nu} = \left(
  \begin{array}{cccc}
    m_{1D} & 0 & 0 \\
    0 & m_{2D} & 0 \\
    0 & 0 & m_{3D} \\
  \end{array}
\right) \, ,
\ee
where 
\be
\label{Unu}
U_{\nu}=  \frac{1}{\sqrt{3}}
\left(
  \begin{array}{cccc}
  1 & 1 & 1 \\
  1 & \omega^2 & \omega \\
  1 & \omega & \omega^2 \\
  \end{array}
\right) \, .
\ee
Notice that our $U_{\nu}$ defined in Eq.~(\ref{Unu}) is just $U_{\nu}= U_{CW}^{\dagger}$. 
At this point, we would like to establish our notations for what will follow. Notice that, in general, a mass matrix is diagonalized by two unitary matrices $U_L$ and $U_R$ i.e. 
\be
\label{diag2}
U_L^{\dagger} \mathcal{M} U_R = \mathcal{M}_D \, ,
\ee
where $M_D$ is a diagonal mass matrix. A mass term of the form $\bar{f}_L^0 \mathcal{M}  f_R^0$ can be rewritten as  $\bar{f}_L^0 U_L U_L^{\dagger} \mathcal{M}   U_R  U_R^{\dagger} f_R^0= \bar{f}_L \mathcal{M}_D f_R$ where $\bar{f}_L^0 U_L = \bar{f}_L$ and $U_R^{\dagger} f_R^0=f_R$.

From Eq.~(\ref{diagonal}), it is clear that
\be
\label{unitary}
U_{\nu L} = U_{\nu R} = U_{\nu} \,.
\ee
A remark is in order at this point. As we will see below, $U_{PMNS}$ is defined as $U_{PMNS}=  U_{\nu L}^{\dagger} U_{l L}= U_{\nu}^{\dagger} U_{l L}$. What $U_{lL}$ might be will be the subject of the section on the charged lepton mass matrix.

The neutrino Dirac masses are
\bea
\label{masses}
m_{1D}&=&  g_{0S}v_0+g_{1S}v+g_{2S}v \\
m_{2D} &=&g_{0S}v_0+g_{1S}v\omega^2+g_{2S}v\omega \\
m_{3D} &=& g_{0S}v_0+g_{1S}v\omega+g_{2S}v\omega^2 
\eea

Reality of the masses require that
\be
\label{real}
g_{2S} = g_{1S}^{*} \, ,
\ee
where we have used $\omega^2=\omega^{\ast}$. Making use of $1+ \omega + \omega^2=0$, $\omega^3=1$ and Eq.~(\ref{real}), we obtain the following sum rules
\be
\label{sumrule1}
m_{1D} + m_{2D} + m_{3D} = 3 g_{0S}v_0 \,.
\ee
\be
\label{sumrule1}
m_{1D}^2 + m_{2D}^2 + m_{3D}^2 = 3 g_{0S}^2v_0^2 + 6 |g_{1S}|^2 v^2  \,.
\ee

%\underline{Note:} $\nu_e$, $\nu_\mu$, $\nu_\tau$ are combinations of $\nu_1$, $\nu_2$, $\nu_3$!\\
%-----------------------------------------------------------------------------
\subsection{Neutrino Majorana mass matrix}
From the Lagrangian
\be
L_M = g_M \: (l_{iR}^{M,T} \: \sigma_2) (i \: \tau_2 \: \widetilde{\chi}) \: l_{jR}^M + H.c.
\ee
In order to make the Lagrangian invariant under $A_4$, we need $\widetilde{\chi}$ to transform as \underline{1} or \underline{3}. For reasons outlined in \cite{hung3} having to do with the constraints coming from the presently known properties of the 125-GeV SM-like boson, it is preferable that the Higgs triplet transforms as \underline{1}.  We recall that

%Also $l_{iR}^{M,T}$ and $l_{jR}^M$ are doublets (2) so the Lagrangian is invariant under SU(2).\\
%$2 \otimes 2 = 3 \oplus 1$ where 3 $\rightarrow$ triplet and 1 $\rightarrow$ singlet.\\

%We have a triplet $\widetilde{\chi} = (3, Y/2 = +1)$ $\Rightarrow$ U(1) invariance.
\be
\widetilde{\chi} = \frac{1}{\sqrt{2}} \vec{\tau} \cdot \vec{\chi} =
\left(
  \begin{array}{cc}
  \frac{1}{\sqrt{2}} \chi^+ & \chi^{++} \\
  \chi^0 & -\frac{1}{\sqrt{2}} \chi{+} \\
  \end{array}
\right)
\ee

When $\big \langle \chi^0 \big \rangle = v_M$ one obtains the following right-handed Majorana mass\\
\be
M_R =
\left(
  \begin{array}{ccc}
  g_{M} \big \langle \chi^0 \big \rangle & 0 & 0 \\
  0 & g_{M} \big \langle \chi^0 \big \rangle & 0 \\
  0 & 0 & g_{M} \big \langle \chi^0 \big \rangle \\
  \end{array}
\right) = g_M v_M \mathbb{I}
\ee

Therefore, the neutrino mass matrix is\\
\be
M_\nu =
\left(
  \begin{array}{cc}
  0 & M_\nu^D \\
  M_\nu^D & M_R \\
   \end{array}
\right)
\ee
Here the $3 \times 3$ see-saw mass matrix for the light neutrinos $(\nu_e, \nu_\mu, \nu_\tau)$ becomes
\be
m_{\nu} \sim - M_\nu^D M_R^{-1} M_\nu^{D,T}
\ee

%-----------------------------------------------------------------------------
\section{Toward $U_{PMNS}$ } 
\subsection{The search for $U_{lL}$}
As mentioned above, we define the diagonalization of a mass matrix by Eq.~({\ref{diag2}). The charged current interaction $g \bar{\nu}_L^{0} \gamma^{\mu} l_L^{0} W_{\mu}^{+}$ can be written in terms of mass eigenstates as $g \bar{\nu}_L^{0} U_{\nu L}  U_{\nu L}^{\dagger} \gamma^{\mu}U_{l L}  U_{l L}^{\dagger}l_L^{0} W_{\mu}^{+}= g \bar{\nu}_L U_{PMNS}\gamma^{\mu} l_L W_{\mu}^{+}$, where $\nu_L$ and $l_L$ are mass eigenstates and where
\be
\label{PMNS}
U_{PMNS}=  U_{\nu L}^{\dagger} U_{l L}= U_{\nu}^{\dagger} U_{l L} \,.
\ee
Notice that, by looking at $U_{PMNS}$ as determined from experiment (\ref{exppmns}), one can safely say that $U_{PMNS} \neq U_{\nu}^{\dagger}$. 
One needs $U_{lL}$ to be different from the unit matrix. But could $U_{lL}$ be? What does the Yukawa coupling of the charged leptons to $\Phi_2$ tell us about $U_{lL}$? (There is a coupling between the mirror and SM charged leptons with the Higgs singlets but its contributions to the masses are negligible as shown in \cite{pqnur}. We will ignore this contribution here.)

The SM Yukawa coupling is
\be
\label{yuk2}
L_{Y} = g_l \bar{l}_{L}  \Phi_2 \ e_{R} + H.c.
\ee
where $
\Phi_2 =
\left(
  \begin{array}{c}
  \phi^+ \\
  \phi^0 \\
   \end{array}
\right)
$, $\big \langle \phi^0 \big \rangle = \frac{v_2}{\sqrt{2}}$\\

From Table~(\ref{assignment}), we have the following $A_4$ assignments: $l_{L}  \sim \underline{3};\ e_{R}  \sim \underline{3}; \Phi_2 \sim \underline{1}$. It can be seen that (\ref{yuk2}) is $A_4$-invariant. From the product rule (\ref{product}), one can see that this $A_4$-invariant Yukawa term gives a degenerate spectrum for the charged leptons, namely
\be
\mathcal{M}_l = g_l \frac{v_2}{\sqrt{2}} 
\left(
  \begin{array}{ccc}
  1 & 0 & 0 \\
  0 & 1 & 0 \\
  0 & 0 & 1 \\
  \end{array}
\right)
\ee
With this one would have $U_{lL} = \mathbb{I}$.
This is unacceptable for two reasons: 1) $m_e \ll m_{\mu} < m_{\tau}$; 2) $U_{lL}$ would be a unit matrix and one would obtain $U_{PMNS}=  U_{\nu L}^{\dagger} $ in disagreement with experiment.

It is then clear that the $A_4$ symmetry which is respected by the Yukawa interactions Eq.~(\ref{yukawa}) giving rise to the neutrino Dirac mass matrix has to be broken in the charged lepton sector. In what follows, we will use a phenomenological approach toward this $A_4$ breaking, namely through an $ans\ddot{a}tz$ for $U_{lL}$. 

%------------------------------------------------------------------------
\subsection{$Ans\ddot{a}tz$ for $U_{lL}$}

As discussed above, strict $A_4$ symmetry in the charged lepton sector would imply that $U_{lL} = \mathbb{I}$. We will parametrize the breaking of $A_4$ by assuming a form which deviates from the unit matrix by a small amount and which is unitary. Using $U_{PMNS}$ and $U_{\nu}$, one can then determine $U_{lL}$. As we shall see below, once $U_{lL}$ is known, one can reconstruct $\mathcal{M}_l \mathcal{M}_l^\dagger$. In this sense, our approach is semi phenomenological because we do not use a specific symmetry assumption to construct the charged lepton mass matrix.

We propose the following $ans\ddot{a}tz$ 
%$A_4$ symmetry requires degenerate charged leptons but we do not want this thing to happen. Therefore, in the unitary scenario of $U_l$ we can use Wolfenstein parameters to construct $U_l$.
\be
U_{lL} = 
\left(
  \begin{array}{ccc}
  1-\frac{\lambda_l^2}{2} & \lambda_l & A_l \lambda_l^3(\rho_l - i \eta_l) \\
  -\lambda_l & 1-\frac{\lambda_l^2}{2} & A_l \lambda_l^2 \\
  A_l \lambda_l^3(1 - \rho_l - i \eta_l) & -A_l \lambda_l^2 & 1 \\
  \end{array}
\right)
\label{a}
\ee
where  $A_l,\ \rho_l,\ \eta_l$ are real parameters of O(1) \cite{Wolfenstein:1983yz}. The subscript $l$ indicates that $A, \ \rho, \ \eta$ belong to charged leptons.

We can now constrain $\lambda_l, A_l,\ \rho_l,\ \eta_l$ based on experimental data of $U_{PMNS}$ and unitarity conditions.

We have
\be
U = U_{PMNS} = U_{\nu}^\dagger U_{lL} = \frac{1}{\sqrt{3}} 
\left(
  \begin{array}{ccc}
  1 & 1 & 1 \\
  1 & \omega^{2*} & \omega^{*} \\
  1 & \omega^{*} & \omega^{2*} \\
  \end{array}
\right) 
\left(
  \begin{array}{ccc}
  1-\frac{\lambda_l^2}{2} & \lambda_l & A_l \lambda_l^3(\rho_l - i \eta_l) \\
  -\lambda_l & 1-\frac{\lambda_l^2}{2} & A_l \lambda_l^2 \\
  A_l \lambda_l^3(1 - \rho_l - i \eta_l) & -A_l \lambda_l^2 & 1 \\  \end{array}
\right)
\nonumber
\ee
Recall that $\omega = e^{i 2 \pi/3}$ so $\omega^* = \omega^2$ and $\omega^{2*} = \omega$. Therefore,
\bea
\label{PMNS2}
U &=& \frac{1}{\sqrt{3}} 
\left(
  \begin{array}{ccc}
  1 & 1 & 1 \\
  1 & \omega & \omega^2 \\
  1 & \omega^2 & \omega \\
  \end{array}
\right) 
\left(
  \begin{array}{ccc}
  1-\frac{\lambda_l^2}{2} & \lambda_l & A_l \lambda_l^3(\rho_l - i \eta_l) \\
  -\lambda_l & 1-\frac{\lambda_l^2}{2} & A_l \lambda_l^2 \\
  A_l \lambda_l^3(1 - \rho_l - i \eta_l) & -A_l \lambda_l^2 & 1 \\
  \end{array}
\right)\\ [0.7ex] \nn
= &\frac{1}{\sqrt{3}}&
\left(
  \begin{array}{ccc}
  A_l \lambda_l^3(1 - \rho_l - i \eta_l) - \frac{\lambda_l^2}{2} - \lambda_l + 1 &  - \left(A_l+\frac{1}{2}\right) \lambda_l^2 + \lambda_l + 1 &  A_l \lambda_l^3(\rho_l - i \eta_l) + A_l \lambda_l^2 + 1 \\[0.7ex]
    \omega^2 A_l \lambda_l^3(1 - \rho_l - i \eta_l) - \frac{\lambda_l^2}{2} - \omega \lambda_l + 1 &  - \left(\omega^2 A_l+\frac{\omega}{2}\right) \lambda_l^2 + \lambda_l + \omega &  A_l \lambda_l^3(\rho_l - i \eta_l) + \omega A_l \lambda_l^2 + \omega^2 \\[0.7ex]
   \omega A_l \lambda_l^3(1 - \rho_l - i \eta_l) - \frac{\lambda_l^2}{2} - \omega^2 \lambda_l + 1 &  - \left(\omega A_l+\frac{\omega^2}{2}\right) \lambda_l^2 + \lambda_l + \omega^2 &  A_l \lambda_l^3(\rho_l - i \eta_l) + \omega^2 A_l \lambda_l^2 + \omega
  \end{array}
\right) 
\eea
Recall the standard parametrization of PMNS matrix (\cite{pdg}, \cite{FritXing})
\bea
U &=& V \ 
\left(
  \begin{array}{ccc}
  1 & 0 & 0 \\
  0 & e^{i \frac{\alpha_{21}}{2}} & 0 \\
  0 & 0 & e^{i \frac{\alpha_{31}}{2}} \\
  \end{array}
\right) \\[0.5ex] 
V &=& \left(
  \begin{array}{ccc}
  c_{12} \ c_{13} & s_{12} \ c_{13} & s_{13} \ e^{-i \delta} \\
  -s_{12} \ c_{23}-c_{12} \ s_{23} \ s_{13} \ e^{i\delta} & c_{12} \ c_{23}-s_{12} \ s_{23} \ s_{13} \ e^{i\delta} & s_{23} \ c_{13} \\
  s_{12} \ s_{23}-c_{12} \ c_{23} \ s_{13} \ e^{i\delta} & -c_{12} \ s_{23}-s_{12} \ c_{23} \ s_{13} \ e^{i\delta} & c_{23} \ c_{13} \\
  \end{array}
\right)
\eea
where $s_{ij} \ \equiv \ sin( \theta_{ij})$, $c_{ij} \ \equiv \ cos (\theta_{ij})$, $\theta_{ij} \ = \ \big[ 0 \ , \ \frac{\pi}{2} \big]$ \\

For the purposes of this paper, the Majorana phases will not be taken into account, i.e. we can set these phases to be equal to zero. Therefore, our PMNS matrix really has the form of V which contains the Dirac phase. 

Let us compare Eq.~(\ref{PMNS2}) with experimental data \cite{Zhao:2012sqa}
\be
\lvert U \rvert = 
\left(
  \begin{array}{ccc}
  0.779...0.848 & 0.510...0.604 & 0.122...0.190 \\
  0.183...0.568 & 0.385...0.728 & 0.613...0.794 \\
  0.200...0.576 & 0.408...0.742 & 0.589...0.775 \\
  \end{array}
\right)
\ee
we have the following constraints
\bea
\label{constraints0}
(i) \quad 0.779 &<& \frac{1}{\sqrt{3}} \lvert A_l \lambda_l^3(1 - \rho_l - i \eta_l) - \frac{\lambda_l^2}{2} - \lambda_l + 1 \rvert < 0.848 \\ \nn
(ii) \quad 0.510 &<& \frac{1}{\sqrt{3}} \lvert - \left(A_l+\frac{1}{2}\right) \lambda_l^2 + \lambda_l + 1 \rvert < 0.604 \\ \nn
(iii) \quad 0.122 &<& \frac{1}{\sqrt{3}} \lvert A_l \lambda_l^3(\rho_l - i \eta_l) + A_l \lambda_l^2 + 1 \rvert < 0.190 \\ \nn
(iv) \quad 0.183 &<& \frac{1}{\sqrt{3}} \lvert \omega^2 A_l \lambda_l^3(1 - \rho_l - i \eta_l) - \frac{\lambda_l^2}{2} - \omega \lambda_l + 1 \rvert < 0.568 \\ \nn
(v) \quad 0.385 &<& \frac{1}{\sqrt{3}} \lvert - \left(\omega^2 A_l+\frac{\omega}{2}\right) \lambda_l^2 + \lambda_l + \omega \rvert < 0.728 \\ \nn
(vi) \quad 0.613 &<& \frac{1}{\sqrt{3}} \lvert A_l \lambda_l^3(\rho_l - i \eta_l) + \omega A_l \lambda_l^2 + \omega^2 \rvert < 0.794 \\ \nn
(vii) \quad 0.200 &<& \frac{1}{\sqrt{3}} \lvert \omega A_l \lambda_l^3(1 - \rho_l - i \eta_l) - \frac{\lambda_l^2}{2} - \omega^2 \lambda_l + 1 \rvert < 0.576 \\ \nn
(viii) \quad 0.408 &<& \frac{1}{\sqrt{3}} \lvert - \left(\omega A_l+\frac{\omega^2}{2}\right) \lambda_l^2 + \lambda_l + \omega^2 \rvert < 0.742 \\ \nn
(ix) \quad 0.589 &<& \frac{1}{\sqrt{3}} \lvert A_l \lambda_l^3(\rho_l - i \eta_l) + \omega^2 A_l \lambda_l^2 + \omega \rvert < 0.775 
\eea
Solving these equations up to $O(\lambda^2)$ we get
\bea
\label{constraints}
-4.8517 &<& A_l < -4.4580 \\ \nn
-0.2404 &<& \lambda_l < -0.1882 \\ \nn
-5.6339 &<& \rho_l < -5.5712 \\ \nn
-4.7160 &<& \eta_l < 4.8912 
\eea
\section{Toward $\mathcal{M}_l {\mathcal{M}_l}^\dagger $}

The knowledge of $U_{lL}$ alone does not allow us to determine the charged lepton mass matrix $\mathcal{M}_l$ for we need also $U_{lR}$. 
On the other hand, we can use $U_{lL}$ to diagonalize $\mathcal{M}_l {\mathcal{M}_l}^\dagger$ as follows.\\
\be
{U_{lL}}^\dagger \mathcal{M}_l {\mathcal{M}_l}^\dagger U_{lL} = 
\left(
  \begin{array}{ccc}
  {m_e}^2 & 0 & 0 \\
  0 & {m_\mu}^2 & 0 \\
  0 & 0 & {m_\tau}^2 \\
  \end{array}
\right)
\ee
giving
\be
\mathcal{M}_l {\mathcal{M}_l}^\dagger = U_{lL} \ .
\left(
  \begin{array}{ccc}
  {m_e}^2 & 0 & 0 \\
  0 & {m_\mu}^2 & 0 \\
  0 & 0 & {m_\tau}^2 \\
  \end{array}
\right) . \ {U_{lL}}^\dagger
\ee
Up to the order of $\lambda^2$ we can approximate $\mathcal{M}_l {\mathcal{M}_l}^\dagger$ to be of the form
\be 
\mathcal{M}_l {\mathcal{M}_l}^\dagger=
\left(
  \begin{array}{ccc}
  (1-\lambda_l^2) \ {m_e}^2 + \lambda_l^2 \  {m_\mu}^2 & \lambda_l ({m_\mu}^2-{m_e}^2) & 0\\
\lambda_l ({m_\mu}^2-{m_e}^2) & \lambda_l^2 \ {m_e}^2+(1-\lambda_l^2) \ {m_\mu}^2 & A_l \lambda_l^2({m_\tau}^2-{m_\mu}^2) \\
0 & A_l \lambda_l^2({m_\tau}^2-{m_\mu}^2) & {m_\tau}^2 \\
  \end{array}
\right)
\label{eqmm}
\ee
\vspace{0.2cm}
From Eq.~(\ref{eqmm}), we can see that $\mathcal{M}_l {\mathcal{M}_l}^\dagger$ is {\em determined completely} by the experimental values of $m_e$, $m_\mu$, $m_\tau$, $\lambda_l$ and $A_l$. Notice that, in the degenerate case $m_e=m_\mu=m_\tau=m$, $\mathcal{M}_l {\mathcal{M}_l}^\dagger$ is reduced to a diagonal matrix $\mathcal{M}_l {\mathcal{M}_l}^\dagger= m^2 \mathbb{I}$ as one should expect. 

A few remarks are in order here. One can view Eq.~(\ref{eqmm}) as a constraint equation on the charged lepton mass matrix $\mathcal{M}_l$. This constraint equation on $\mathcal{M}_l {\mathcal{M}_l}^\dagger$ satisfies the experimental constraints on $U_{PMNS} $ as long as $\lambda_l$ and $A_l$ are within the allowed ranges (\ref{constraints}). To be able to determine the form of $\mathcal{M}_l$, it is clear that one has to impose some kind of symmetry or at the very least make an  
$ans\ddot{a}tz$ on $\mathcal{M}_l$ itself as long as $\mathcal{M}_l {\mathcal{M}_l}^\dagger$ satisfies Eq.~(\ref{eqmm}).

Based on the above discussion, it is tempting to propose a similar $ans\ddot{a}tz$ for the quark sector for the following reason. The charged leptons as well as the quarks obtain their masses through the couplings with the Higgs doublet $\Phi_2$. It might not be unreasonable to speculate that whatever mechanism giving rise to mass mixings in mass matrices could be similar for both quarks and charged leptons. One might have
\bea
\label{Ud}
U_{lL} &\rightarrow& U_{dL}  \\ \nn
\lambda_l, A_l,\ \rho_l,\ \eta_l &\rightarrow& \lambda_d, A_d,\ \rho_d,\ \eta_d
\eea
\bea
\label{Uu}
U_{lL} &\rightarrow& U_{uL}  \\ \nn
\lambda_l, A_l,\ \rho_l,\ \eta_l &\rightarrow& \lambda_u, A_u,\ \rho_u,\ \eta_u
\eea
With the knowledge of $V_{CKM} = U_{uL}^{\dagger} U_{dL}$ \cite{CKM}, one can constraint the above parameters. Furthermore, $\mathcal{M}_u {\mathcal{M}_u}^\dagger$ and $\mathcal{M}_d {\mathcal{M}_d}^\dagger$ could have similar forms to the right-hand side of Eq.~(\ref{eqmm}) with the replacements $\lambda_l, A_l \rightarrow \lambda_u, A_u$, $m_e, m_\mu, m_\tau \rightarrow m_u, m_c, m_t$ and $\lambda_l, A_l \rightarrow \lambda_d, A_d$, $m_e, m_\mu, m_\tau \rightarrow m_d, m_s, m_b$ respectively. It is beyond the scope of this paper to go into details of this possibility. This will be treated elsewhere.

One important remark is in order here. Since the sources of masses for the neutrinos (Higgs singlets and triplet) and for the charged leptons and quarks (Higgs doublet) are very different from each other, it might not be surprising, within the context of this paper, that $U_{PMNS}$ is {\em very different} from $V_{CKM}$.

%----------------------------------------

%====================
\section{Conclusion} 
We have presented in this manuscript a model of neutrino masses and mixings based on the discrete symmetry group $A_4$ as applied to the electroweak(EW)-scale right-handed neutrino model of \cite{pqnur}. In particular, this $A_4$ symmetry is applied to the Higgs singlets which are responsible for the neutrino Dirac masses of the EW-scale $\nu_R$ model with the aim of obtaining a particular form of matrix namely $U_{CW}$ (Eq.~(\ref{CW})), which plays a crucial role in $U_{PMNS}$. The Higgs singlet was introduced in \cite{pqnur} in order to give the Dirac part of the neutrino masses. By applying the $A_4$ symmetry to this sector, we found that the Higgs singlet is increased from {\em one} in the original model to {\em four} i.e. \underline{1} + \underline{3} of $A_4$. The diagonalization of the neutrino Dirac mass matrix generated by the Yukawa coupling of the left-handed doublets $(\nu_L, e_L) \sim \underline{3}$, the right-handed doublets $(\nu_R, e^{M}_R) \sim \underline{3}$ with these four Higgs singlets is found to be realized by the matrix $U_{\nu} =U_{CW}^{\dagger}$ (Eq.~(\ref{Unu})). This is in contrast with many popular $A_4$-based models where this type of matrix  is the one that diagonalizes the {\em charged lepton mass matrix}. This is our first step in getting to $U_{PMNS}$ namely $U_{PMNS}= U_{\nu}^{\dagger} U_{lL}$. In obtaining $U_{\nu}$, we also derive a couple of sum rules concerning the Dirac masses of the neutrinos. These might turn out to be useful in future studies of neutrino oscillations.

One particular interesting feature of this scheme is the fact that $U_{\nu}$ is generated by the Higgs singlets which do not affect the known properties of the newly discovered 125-GeV SM-like scalar \cite{hung3}. Notice that scenarios involving more than two Higgs doublets might encounter very very tight constraints which may be hard to satisfy.

 The second piece of $U_{PMNS}$, namely $U_{lL}$, comes from the breaking of the $A_4$ symmetry in the charged lepton sector as we have shown. It is proportional to the unit matrix in the exact symmetry case (degenerate charged leptons). We take a phenomenological approach by parametrizing the deviation from the unit matrix in terms of a Wolfenstein-like unitary matrix (Eq.~(\ref{a})). We obtain constraints on the parameters of that matrix by using the experimental values of $U_{PMNS}$. Since $U_{lL}$ diagonalizes the lepton mass matrix "squared", namely ${U_{lL}}^\dagger \mathcal{M}_l {\mathcal{M}_l}^\dagger U_{lL} $, we obtain an equation for $\mathcal{M}_l {\mathcal{M}_l}^\dagger$ (Eq.~(\ref{eqmm})) whose right-hand side is determined entirely by experimental values of the charged lepton masses and the phenomenologically extracted parameters of $U_{lL}$.
 
 As shown in \cite{pqnur} and in this manuscript, the sources of masses for the neutrinos and for the charged leptons are entirely different from each other: Higgs singlets and triplet for the neutrinos and Higgs doublet for the charged leptons. Since the quarks also obtain their masses from the Higgs doublet and since $V_{CKM}$ deviates a "little" from the unit matrix, we postulate that $U_{uL}$ and $U_{dL}$ which appear in $V_{CKM} = U_{uL}^{\dagger} U_{dL}$ have the same form as $U_{lL}$ but endowed with their own parameters. In this context, it is very appealing to see why $U_{PMNS}$ is {\em very different} from $V_{CKM}$.
 
\section{Acknowledgements}
This work was supported by US DOE grant DE-FG02-97ER41027.

%
%
%%%%%%%%%%%%%%%%%%%%
\appendix
%%%%%%%%%%%%%%%%%%%%
%
%
%%%%%%%%%%%%%%%%%%%%	
%%%%%%%%%%%%%%%%%%%%
%\FloatBarrier
%%%%%%%%%%%%%%%%%%%%
%
%
%
%%%%%%%%%%%%%%%%%%%%
%

\end{document}